\documentclass{PoS}

\usepackage{aas_macros}
\usepackage{color}
\usepackage{multirow}
\usepackage{url}

\title{Mapping dark matter in the Milky Way, a synopsis}

\ShortTitle{Mapping dark matter in the Milky Way, a synopsis}

\author{Miguel Pato\footnote{Speaker.}\\
        The Oskar Klein Centre for Cosmoparticle Physics, \\Department of Physics, Stockholm University,\\ AlbaNova, SE-106 91 Stockholm, Sweden\\
        E-mail: \email{migpato@gmail.com}}

\author{Fabio Iocco\\
        ICTP South American Institute for Fundamental Research, and\\ Instituto de F\'isica Te\'orica - Universidade Estadual Paulista (UNESP),\\ Rua Dr. Bento Teobaldo Ferraz 271, 01140-070 S\~{a}o Paulo, SP Brazil\\
        E-mail: \email{iocco@ift.unesp.br}}

\abstract{Mapping the dark matter distribution across our Galaxy represents a central challenge for the near future as a new generation of space-borne and ground-based astronomical surveys swiftly comes online. Here we present a synopsis of the present status of the field, reviewing briefly the baryonic content and the kinematics of the Milky Way and outlining the methods used to infer the dark matter component. The discussion then proceeds with some of the latest developments based on our own work. In particular, we present a new compilation of kinematic measurements tracing the rotation curve of the Galaxy and an exhaustive array of observation-based baryonic models setting the contribution of stellar bulge, stellar disc and gas to the total gravitational potential. The discrepancy between these two components is then quantified to derive the latest constraints on the dark matter distribution and on modified Newtonian dynamics. We shall end with an overview of future directions to improve our mapping of the dark matter distribution in the Milky Way.}

\FullConference{The 34th International Cosmic Ray Conference,\\
		30 July- 6 August, 2015\\
		The Hague, The Netherlands}

\begin{document}

\section{Preamble}\label{sec:intro}
\par It is almost one hundred years since the presence of an invisible component of matter across the cosmos has first been suggested. The earliest remarks on the dynamical contribution of an invisible component date back to the seminal works of Kapteyn \cite{Kapteyn1922} and Oort \cite{Oort1927} in the 1920s, shortly before Zwicky \cite{Zwicky1933} postulated the existence of large amounts of dark matter to explain the dynamics of the Coma cluster of galaxies. The history of dark matter in individual galaxies started soon afterwards, in 1939, when Babcock \cite{Babcock1939} measured the Doppler shift of emission and absorption lines in the gas of our closest spiral galaxy, Andromeda. Throughout the following decades, several studies (e.g.~\cite{RubinFord1970}) confirmed the fast rotation of Andromeda and indicated an almost flat rotation curve. This same behaviour was then observed in numerous other spiral galaxies throughout the 1970s and 1980s, lending support to the dark matter paradigm. Curiously, the case of our Galaxy, a spiral itself, is much more complicated. While many external galaxies happen to lie towards particularly convenient lines of sight with particularly convenient inclination angles, in the Milky Way there is not much we can do: our inside position makes it extremely hard to measure accurate distances and velocities. Therefore, determining the rotation curve of our Galaxy remains a challenging enterprise still today. Current and forthcoming astronomical surveys, however, hold the promise of staggering improvements in the field over the coming years.

\par Here we give a synopsis of the ongoing efforts in determining the dark matter distribution across the Milky Way. We start with a brief tour of the Galaxy in Sec.~\ref{sec:tour}, separated into the different baryonic components and the total mass distribution. Sec.~\ref{sec:dm} is instead devoted to the determination of the dark matter content, including a description of the so-called local and global methods as well as their strengths and pitfalls. We then summarise the latest developments based on our recent works \cite{2015NatPh..11..245I,2015arXiv150406324P,2015ApJ...803L...3P,Iocco:2015iia} in Sec.~\ref{sec:latest}, before finalising in Sec.~\ref{sec:future} with an outlook on the future prospects in the field. These proceedings are intended as a synopsis only, not as a comprehensive review, for which we refer the interested reader to the list of references, in particular to the books \cite{2008gady.book.....B,1998gaas.book.....B} and the recent review \cite{Read2014}.

\section{Tour of the Galaxy}\label{sec:tour}
\par The Milky Way is a complex, gravitationally-bound system of stars, gas and dark matter. In the very centre of the Galaxy sits a supermassive black hole with a mass $4.4\times 10^6\,$M$_{\odot}$, as inferred from the orbits of tens of S-stars in the inner $0.01\,$pc \cite{Gillessen2009,Genzel2010}. The gravitational influence of the supermassive black hole dominates the central pc, but becomes negligible in the regions we are interested in, namely above $1\,$kpc off the Galactic centre, so we shall neglect it in the following. For our purposes here, there are three main baryonic components of the Galaxy: \emph{stellar bulge}, dominating the innermost $2-3\,$kpc with its barred shape and total mass of order $10^{10}\,$M$_{\odot}$; \emph{stellar disc}, often split in thin and thick populations with a fast decaying radial profile and total mass around $(2-5)\times10^{10}\,$M$_{\odot}$; and \emph{gas}, comprising about $5\times10^{9}\,$M$_{\odot}$ of molecular, atomic and ionised hydrogen (and heavier elements). The stellar disc and the gas have a marked spiral structure with four main arms (see Ref.~\cite{Vallee2014} and references therein), where the Sun lies in between the Perseus and Sagittarius arms, in a local armlet, at a distance to the Galactic centre $R_0=7.5-8.5\,$kpc \cite{Gillessen2009,Ando2011,Malkin2012,Reid2014}. The local standard of rest (LSR) moves around the Galactic centre at the so-called local circular velocity $v_0=210-250\,$km/s \cite{ReidBrunthaler2004,Reid2009,Bovy2009, McMillan2010, Bovy2012,Reid2014}, while the Sun motion with respect to the LSR reads $(U,V,W)_{\odot}=(10-11,\,5-26,\,7-9)\,$km/s \cite{DehnenBinney1998,Schoenrich2010,Bovy2012,Reid2014}. The three baryonic components described above are supposed to be embedded in a \emph{dark matter halo}, likely extending hundreds of kpc but whose properties are not well constrained at present. To sum up, the total gravitational potential of our Galaxy receives contributions from the baryons (bulge, disc, gas) and dark matter separately:
\begin{equation}\label{eq:phi}
\phi_{\textrm{tot}} = \phi_{\textrm{bulge}} + \phi_{\textrm{disc}} + \phi_{\textrm{gas}} + \phi_{\textrm{dm}} \, . 
\end{equation}
The key challenge is then to constrain the different terms in this equation. Photometric data trace the individual baryonic components $\phi_{\textrm{bulge}}$, $\phi_{\textrm{disc}}$, $\phi_{\textrm{gas}}$, while kinematic data trace the total gravitational potential $\phi_{\textrm{tot}}$. The dark matter contribution follows from the comparison of these two inputs.

\subsection{Baryonic components (photometry)}\label{sec:baryon}
\par Photometric data play a crucial role in tracing and discerning the different baryonic components. In the following we go through the specifics of each component and divide the discussion into morphology and normalisation. Tab.~\ref{tab:models} (adapted from Ref.~\cite{2015arXiv150406324P}) summarises the different baryonic models used for the morphology of each component.

\begin{table}
\begin{center}
\begin{tabular}{ |l| ccc l c l c c|	} 
%\hline
%\hline
\cline{2-9}
 \multicolumn{1}{c|}{} &	 & model & & specification && data && Ref.\\
\hline
\multirow{7}{*}{bulge} 	& & 1 & & 				exponential E2			&& optical		&& \cite{Stanek1997}\\
			& & 2 & & 				gaussian G2 			&& optical		&& \cite{Stanek1997}\\
			& & 3 & & 				gaussian plus nucleus 		&& infrared		&& \cite{Zhao1996}\\
			& & 4 & & 				truncated power law		&& infrared		&& \cite{BissantzGerhard2002}\\
			& & 5 & & 				power law plus long bar		&& optical | infrared	&& \cite{LopezCorredoira2007}\\
			& & 6 & & 				truncated power law		&& optical | infrared	&& \cite{Vanhollebeke2009}\\
			& & 7 & & 				double ellipsoid 		&& infrared		&& \cite{Robin2012}\\
\hline
\multirow{5}{*}{disc} 	& & 1 & & 				thin plus thick 		&& optical		&& \cite{HanGould2003}\\
			& & 2 & & 				thin plus thick 		&& optical		&& \cite{CalchiNovatiMancini2011}\\
			& & 3 & & 				thin plus thick plus halo	&& optical		&& \cite{deJong2010}\\
			& & 4 & & 				thin plus thick plus halo	&& optical		&& \cite{Juric2008}\\
			& & 5 & & 				single maximal disc		&& optical		&& \cite{Bovy:2013raa}\\
\hline
\multirow{2}{*}{gas} 	& & 1 & & 				H$_2$, HI, HII			&& optical | microwave | radio		&& \cite{Ferriere1998}\\
			& & 2 & & 				H$_2$, HI, HII 			&& optical | microwave | radio		&& \cite{Moskalenko2002}\\	
%\hline
\hline
\end{tabular}
\caption{Details of the baryonic models used to set the stellar bulge, stellar disc and gas in our Galaxy. Table adapted from Ref.~\cite{2015arXiv150406324P}; please refer to Refs.~\cite{2015NatPh..11..245I,2015arXiv150406324P} for a full description of each model.}\label{tab:models}
\end{center}
\end{table}

\paragraph{Stellar bulge}
The morphology of the bulge has been extensively mapped with the help of optical and infrared surveys (such as COBE/DIRBE, OGLE and 2MASS), revealing an elongated triaxial structure, known as the bar, with near end at positive Galactic longitudes. The inclination of the bar with respect to the Galactic centre line of sight has not been precisely measured and angles reported in the literature \cite{Stanek1997,Zhao1996,BissantzGerhard2002,LopezCorredoira2007,Vanhollebeke2009,Robin2012} vary between $10^{\circ}$ and $45^{\circ}$. Similarly, the spatial distribution has been fitted with several variants of exponential, gaussian and power law profiles (cf.~Tab.~\ref{tab:models}). Regarding the normalisation of the bulge, one possibility is to use the measurement of the microlensing optical depth towards the central Galactic region \cite{MACHO2005,OGLEII2006,EROS22006}. This procedure is particularly convenient in deriving model-independent constraints since the optical depth does not depend on the individual masses of the lenses but only on their mass density along the line of sight \cite{1986ApJ...304....1P,1997ApJ...486L..19K,2011JCAP...11..029I}.

\paragraph{Stellar disc(s)}
Optical photometric surveys, most prominently SDSS, have been instrumental in pinpointing the structure of the stellar disc. Based on these data, several works in the literature \cite{HanGould2003,CalchiNovatiMancini2011,deJong2010,Juric2008} describe the morphology of the disc with a double exponential profile \cite{2008gady.book.....B} and two components, a thin population (of scale height $\sim 0.25\,$kpc) and a thick population (of scale height $\sim 0.75\,$kpc). There are also works implementing a single effective component \cite{Bovy:2013raa}. The normalisation of the disc, in particular the local total stellar surface density, can be pinned down with star censuses \cite{2006MNRAS.372.1149F,2012ApJ...751..131B} or dynamically with a Jeans analysis of the kinematics of specific tracer stars \cite{Bovy:2013raa}.

\paragraph{Gas}
The gas in our Galaxy is mainly composed of molecular, atomic and ionised hydrogen (H$_2$, HI and HII, respectively). Each component is probed with different observations: CO lines for molecular gas; $21\,$cm line for atomic gas; H$\alpha$ line and dispersion measures of pulsars for ionised gas. The observations show a very patchy morphology in the very inner $10\,$pc of the Galaxy \cite{Ferriere2012} and a disc-like structure otherwise, including features such as a central molecular zone and a holed disc in the inner $2\,$kpc \cite{Ferriere2007}. The distribution at larger scales has been compiled in Refs.~\cite{Ferriere1998,Moskalenko2002}. Overall, the gas content is dominated by H$_2$ in the inner Galaxy and HI in the outer Galaxy. The main normalisation uncertainties of the gas component arise from the poorly constrained CO-to-H$_2$ factor \cite{Ferriere1998,Ackermann2012} for H$_2$ and from a factor $\sim 2$ discrepancy between different $21\,$cm line surveys in the inner $15\,$kpc for HI \cite{2008A&A...487..951K}.

\subsection{Total gravitational potential (kinematics)}\label{sec:total}
\par There are numerous kinematic observables used to track down the total mass distribution across the Galaxy. Some include timing arguments in the Local Group \cite{1959ApJ...130..705K} or the kinematics of Milky Way satellites \cite{2003A&A...397..899S}. Here we shall focus instead on rotation curve tracers and star population tracers.

\paragraph{Rotation curve tracers}
These include mainly young objects or regions that track the Galactic rotation. While in external galaxies the only tracer available to us is the gas, in the case of our Galaxy we can also resolve stars and star-forming regions and use those as kinematic tracers. Today, despite the challenges posed by our position inside the Galaxy, the rotation curve can be precisely measured from about 1 to 30 kpc off the Galactic centre using gas \cite{Fich1989,Malhotra1995,McClure-GriffithsDickey2007,HonmaSofue1997,BurtonGordon1978,Clemens1985,Knapp1985,Luna2006,Blitz1979,Fich1989,TurbideMoffat1993,BrandBlitz1993,Hou2009}, stars \cite{FrinchaboyMajewski2008,Durand1998,Pont1994,Pont1997,DemersBattinelli2007,Battinelli2013} and masers in star forming regions \cite{Reid2014,Honma2012,StepanishchevBobylev2011,Xu2013,BobylevBajkova2013}. A new, up-to-date compilation of tracers has been presented in Ref.~\cite{2015NatPh..11..245I} and will be made publicly available soon in Ref.~\cite{IoccoPatoTool}.

\paragraph{Star population tracers}
In a galaxy like our own, star encounters are relatively rare and stars feel on average the smooth gravitational potential of the whole system. We can therefore treat a given set of stars as a collisionless gas and apply the collisionless Boltzmann equation. The first momentum of the Boltzmann equation gives the Jeans equations, which relate the total gravitational potential to the density distribution and velocity dispersions of a given population of selected stars. Star populations are thus invaluable kinematic tracers and have in fact been used to pinpoint the total gravitational potential in the outer halo \cite{2006MNRAS.369.1688D,2008ApJ...684.1143X,2014ApJ...785...63B,2014ApJ...794...59K}, across the disc \cite{Bovy:2013raa,2014ApJ...794..151L} and above and below the Galactic plane \cite{1991ApJ...367L...9K,2004MNRAS.352..440H,2012ApJ...751...30M,2012ApJ...756...89B,2015A&A...573A..91M}.

\section{Dark matter content}\label{sec:dm}
\par As highlighted in Sec.~\ref{sec:tour}, kinematic data fix the total gravitational potential whereas photometric data set the individual baryonic components; the comparison between the two can then be used to infer the dark matter distribution across the Milky Way. There are two classes of methods designed to perform this comparison: \emph{local methods}, which use observations from one specific patch of the sky to derive the dark matter content therein, and \emph{global methods}, which use data from different regions to infer the dark matter content elsewhere. An excellent example of the performance of local and global methods is the measurement of the local dark matter density, see Ref.~\cite{Read2014} for an extensive review. Here we give a brief overview of the strengths and pitfalls of each method in measuring the local dark matter density, and their complementarity.

\subsection{Local methods}\label{sec:local}
\par Local methods couple the Jeans and Poisson equations, thereby relating the total mass density in the solar neighbourhood with the kinematics of a given nearby population of stars. The simplest configuration is the so-called Oort limit \cite{1932BAN.....6..249O,2008gady.book.....B}, in which the mixed radial-vertical term in the Jeans equation is neglected and, assuming a flat rotation curve, the total mass density can be written in terms of a double derivative of the velocity dispersion of the stars in the vertical direction. This simple scheme has been extended over the years to include various corrections with simulations often used as a test bed. A critical step in constraining the dark matter density in the solar neighbourhood is to account for the local dynamical contribution of baryons, which is not precisely known. This, together with other complexities of the method, induces substantial uncertainties precluding a very precise measurement of the local dark matter density, at least with current data. Several works in the literature implement local methods with different approaches, assumptions and data, see e.g.~Refs.~\cite{2011MNRAS.416.2318G,2012ApJ...746..181S, 2012MNRAS.425.1445G, 2013ApJ...772..108Z, Read2014}. The main advantage of this technique is to provide a truly local measurement of the dark matter density in the solar neighbourhood without assuming a global mass model for the whole Galaxy. The most important pitfalls include the difficulty of constructing a clean sample of tracers and of treating kinematically inhomogeneous populations, and the significant uncertainties pertaining the contribution of baryons (most notably, gas) in the solar neighbourhood. For a critical discussion of these pitfalls, see Ref.~\cite{2015arXiv150600384H}.

\subsection{Global methods}\label{sec:global}
\par Global methods are based instead on the mass modelling of the Milky Way as a whole. Typically, mass models for the baryonic components (in particular, stellar bulge and stellar disc) and for the dark matter halo are assigned with a number of free parameters that are then fitted to kinematic data across the Galaxy. The data available are rather broad, ranging from terminal velocities of the gas to the measurement of the Oort constants and the kinematics of stellar objects at large distances from the Galactic centre. This extensive fitting procedure fixes several quantities such as the local dark matter density to a rather good precision, often around 10\% or better. The literature of global methods is rather rich, see e.g.~Refs.~\cite{1998MNRAS.294..429D, 2009PASJ...61..227S, CatenaUllio2010, 2010A&A...509A..25W, 2010A&A...523A..83S, 2011JCAP...11..029I, 2011MNRAS.414.2446M, 2013JCAP...07..016N, 2015arXiv150405368S}. The major advantage of this technique is perhaps the wealth of available data, which are invaluable in reconstructing the overall structure of our Galaxy. The fundamental drawback of global measurements of the local dark matter density is that they are particularly prone to systematics associated, for instance, to baryonic modelling (see e.g.~Ref.~\cite{2015arXiv150406324P}) or to the shape of the dark matter halo (see e.g.~Refs.~\cite{2010PhRvD..82b3531P,2014JCAP...09..004B}).

\vspace{0.5cm}
\par  Generically, the upshot is that local methods are usually less precise but more robust than global methods are, and both bring complementary input regarding the dark matter distribution in the Milky Way. This is rather evident in the case of the local dark matter density: since local methods probe the solar neighbourhood itself and global methods probe a usually spherical shell around the Galactic centre, the comparison between both determinations can be used as a diagnostic test on the shape of the dark matter halo, in particular whether this is oblate or prolate (see Ref.~\cite{Read2014} for a discussion). Given the sizeable uncertainties at present, the outcome of such comparison remains inconclusive.

\section{Latest developments}\label{sec:latest}
\par That dark matter dominates the overall mass budget in the Milky Way has been inferred from kinematic tracers in the outskirts of the Galaxy (e.g.~Ref.~\cite{2008ApJ...684.1143X}). However, closer to the centre, namely inside the solar circle, where baryons give an increasingly important contribution to the total gravitational potential, dark matter constraints are weaker and often lack the level of accuracy required today for particle dark matter searches and for interpreting the results of galaxy formation simulations with the highest resolution. In this context, we have addressed first the presence and second the distribution of dark matter in the innermost regions of the Galaxy (i.e.~inside the solar circle), making use of an extensive array of observation-based descriptions of the different baryonic components (cf.~Sec.~\ref{sec:baryon}) and a comprehensive compilation of rotation curve tracers (cf.~Sec.~\ref{sec:total}). Here we give a brief account of our recent work based on Refs.~\cite{2015NatPh..11..245I,2015arXiv150406324P,2015ApJ...803L...3P,Iocco:2015iia}, to which the reader is referred for a detailed discussion.

\subsection{Evidence for dark matter}\label{sec:evidence}
\par
\begin{figure*}[t]
\includegraphics[width=0.5\textwidth,height=5cm]{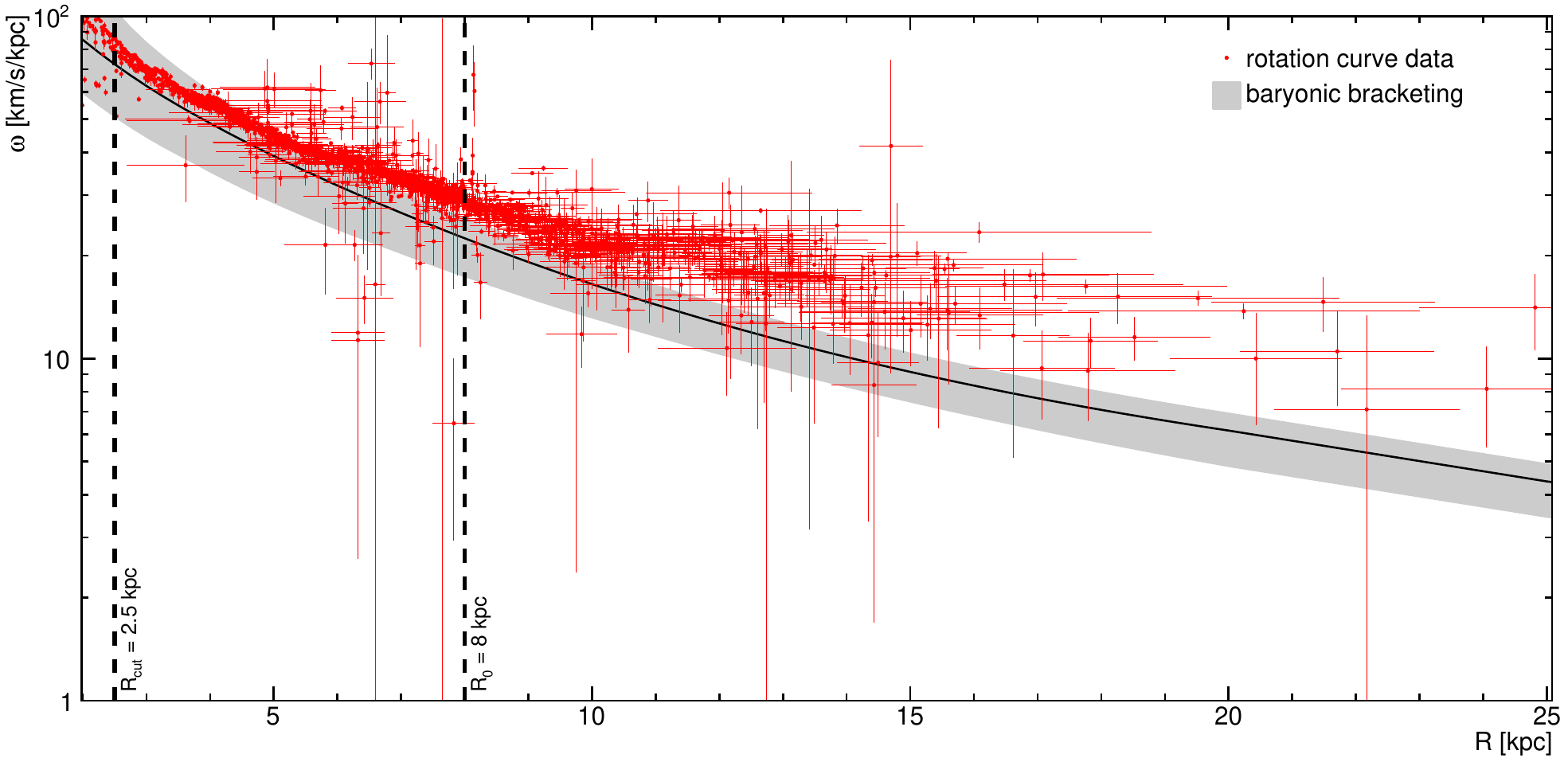}
\includegraphics[width=0.5\textwidth,height=5cm]{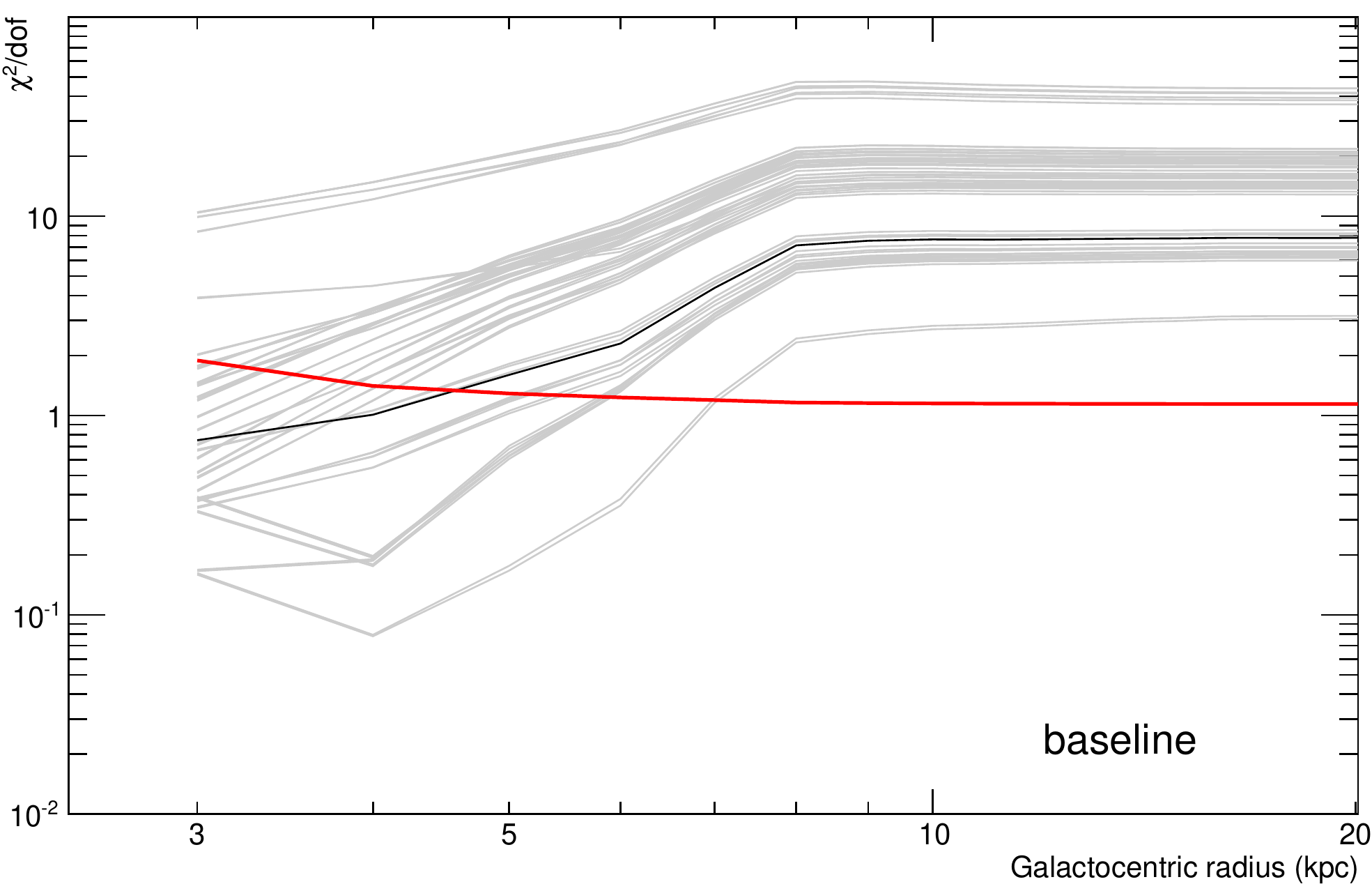}
\caption{Evidence for dark matter. The left panel shows the compilation of rotation curve tracers in red and the convolution of all baryonic models (together with their 1$\sigma$ uncertainties) in grey. The right panel displays the cumulative goodness of fit as a function of Galactocentric radius for each single baryonic model; the red line corresponds to a 5$\sigma$ exclusion. The black line in both panels corresponds to a representative baryonic model. This figure assumes a distance of the Sun to the Galactic centre $R_0=8\,$kpc, a local circular velocity $v_0=230\,$km/s and the peculiar solar motion of Ref.~\cite{Schoenrich2010}. Plots taken from Refs.~\cite{2015NatPh..11..245I,2015arXiv150406324P}. Please refer to Ref.~\cite{2015NatPh..11..245I} for further details and to Ref.~\cite{IoccoPatoTool} for retrieving the kinematic data.}\label{fig:evidence}  
\end{figure*}

 In Ref.~\cite{2015NatPh..11..245I}, we have addressed the very observational evidence for part of the total gravitational potential being generated by a non-visible component. The rationale behind our approach was to compare the rotation curve of the Galaxy with the contribution due to the observed baryonic components in order to quantify the discrepancy between the two as a function of the Galactocentric radius and to assess the impact of all relevant uncertainties. Fig.~\ref{fig:evidence} shows the key results of our analysis. In the left panel, the rotation curve (red data points) is compared directly to the bracketing of the baryonic contribution (grey band). Notice that the baryonic band corresponds to the convolution of the 70 alternative baryonic models that can be generated with the seven bulges, five discs and two gas configurations in Tab.~\ref{tab:models}. The discrepancy between the rotation curve and the envelope representing the baryonic uncertainty appears already by eye; nonetheless, a thorough statistical analysis needs to be performed to properly assess the level of discrepancy. The outcome of such analysis is presented in the right panel of Fig.~\ref{fig:evidence} in terms of the cumulative goodness of fit as a function of the Galactocentric radius for each of the 70 baryonic models. All baryonic models fail to fit the observed rotation curve -- at more than five sigma -- already inside the solar circle. Most of them actually fall short of matching the total gravitational potential already at Galactocentric radii of approximately $5\,$kpc. These results have been shown to be robust against an exhaustive set of uncertainties (including fundamental Galactic parameters, peculiar solar motion, baryonic uncertainties, data selection and systematics, cf.~supplementary information in Ref.~\cite{2015NatPh..11..245I}). It is important to stress at this point that the dark matter hypothesis has not been introduced at all in this analysis, and the method described so far simply quantifies the discrepancy between the total and baryonic gravitational potentials without any theoretical prejudice.

\subsection{Dark matter distribution}\label{sec:distr}
\par Given the evidence that the total gravitational potential of the Milky Way cannot be accounted for by visible matter only, we set out to infer the distribution of dark matter that best accounts for the missing mass. Two different approaches were adopted for this purpose: profile fitting \cite{2015arXiv150406324P} and profile reconstruction \cite{2015ApJ...803L...3P}.

\begin{figure*}[t]
\includegraphics[width=0.5\textwidth,height=5cm]{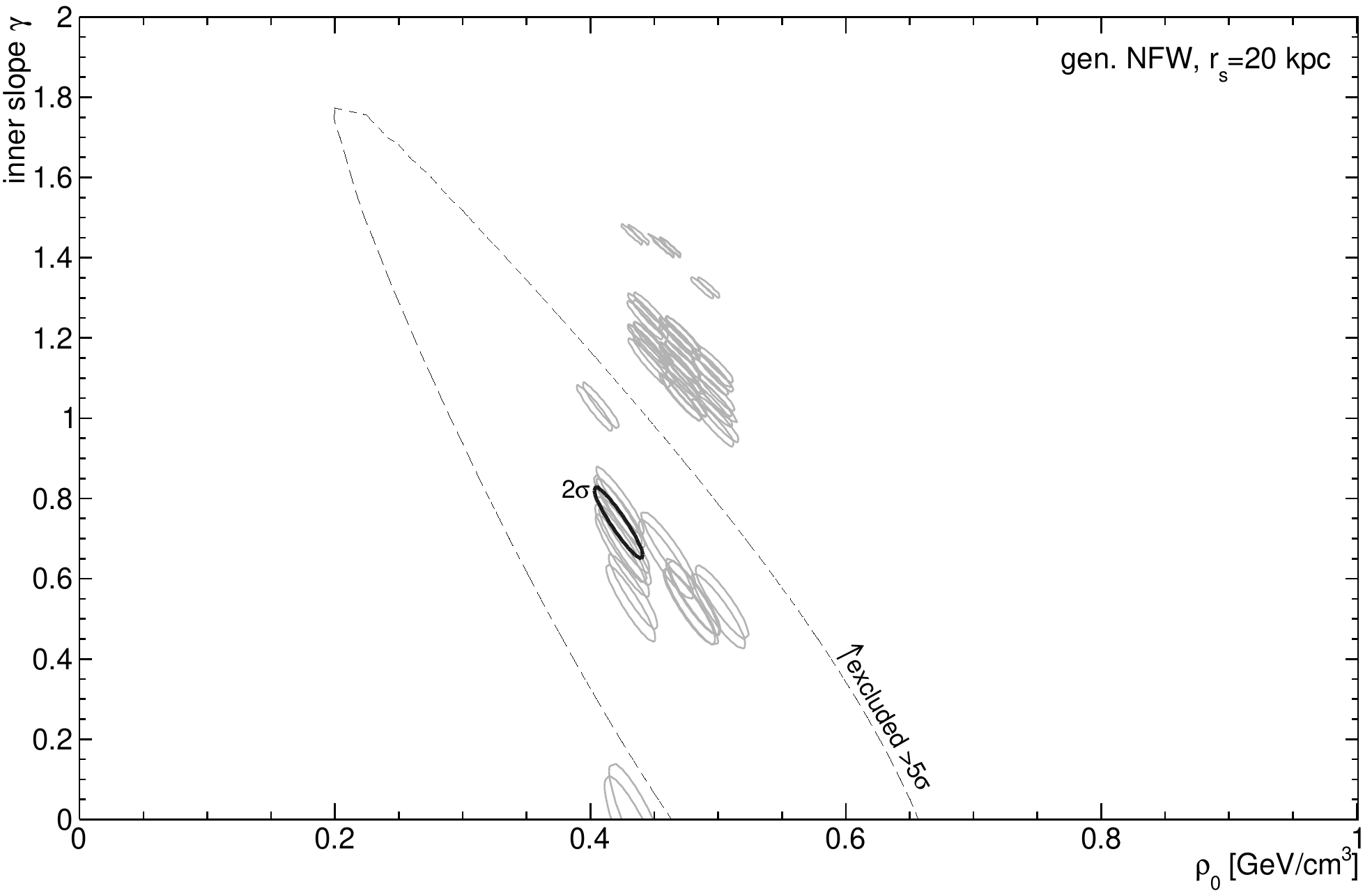}
\includegraphics[width=0.5\textwidth,height=5cm]{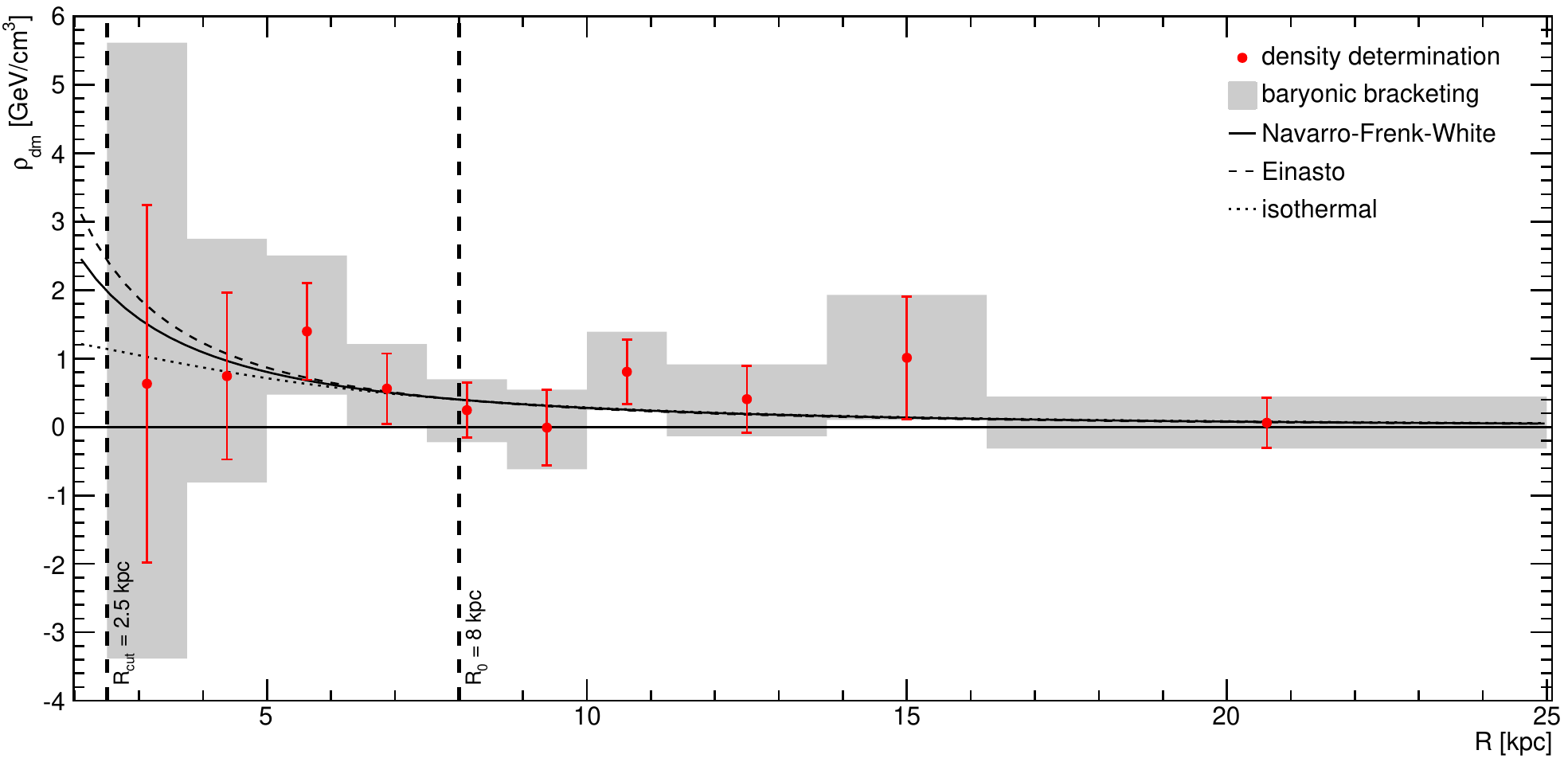}
\caption{Inferring the dark matter distribution. The left panel shows the 2$\sigma$ confidence regions for a generalised NFW profile (with fixed scale radius $r_s=20\,$kpc) and each single baryonic model. The confidence region corresponding to a representative baryonic model is marked in black; for this baryonic model, the 5$\sigma$ goodness-of-fit region is also shown. The right panel displays a non-parametric reconstruction of the dark matter profile, obtained without assuming any functional form. The red error bars correspond to the reconstruction for a representative baryonic model, while the grey band encompasses the reconstruction for all baryonic models. This figure assumes a distance of the Sun to the Galactic centre $R_0=8\,$kpc, a local circular velocity $v_0=230\,$km/s and the peculiar solar motion of Ref.~\cite{Schoenrich2010}. Plots taken from Refs.~\cite{2015arXiv150406324P,2015ApJ...803L...3P}. Please refer to Refs.~\cite{2015arXiv150406324P,2015ApJ...803L...3P} for further details.}\label{fig:density}  
\end{figure*}

\paragraph{Profile fitting}
First, we performed a simple fit to the kinematic data, in which the dark matter distribution is described by a spherical profile with a parameteric, pre-assigned functional form. This sort of global method has been extensively studied in the literature, cf.~Sec.~\ref{sec:global} and references therein. Our analysis in Ref.~\cite{2015arXiv150406324P} (see also Ref.~\cite{2011JCAP...11..029I} for earlier constraints) focussed on the generalised Navarro-Frenk-White (NFW) and Einasto profiles. The left panel of Fig.~\ref{fig:density} reproduces the results for the generalised NFW profile in terms of the local dark matter density $\rho_0$ and the inner slope $\gamma$ for all 70 baryonic models mentioned in Sec.~\ref{sec:evidence}. It is clear that our ignorance about the actual baryonic morphology is preventing a more precise determination of the dark matter profile, especially its slope towards the inner part of the Galaxy. In particular, we find that the local dark matter density is (slightly) degenerate with the morphology of the stellar disc and the inner slope is (heavily) degenerate with the morphology of the stellar bulge. Therefore, in the future a more precise reconstruction of disc and bulge would be extremely valuable in achieving better constraints on the dark matter distribution in the Milky Way.

\paragraph{Profile reconstruction}
The previous technique is sound but relies on the assumption of a given parametric profile: the fitted parameters will be useful only if the original assumption is sufficiently accurate. Therefore, it is important to reconstruct -- with current or upcoming data -- the dark matter profile directly from the data, without any parametric assumption. In Ref.~\cite{2015ApJ...803L...3P}, we have adopted a method to perform such profile reconstruction assuming a spherical dark matter distribution. Specifically, we have used the magnitude and slope of the residuals between the rotation curve and the baryonic contribution to reconstruct the dark matter density in the inner $25\,$kpc of the Milky Way. The results are shown in the right panel of Fig.~\ref{fig:density}. The dark matter density profile obtained with this method is not very precise and in fact it is not possible to discriminate between different profiles with current data. Notice that this is entirely consistent with the evidence for a dark matter component described in Sec.~\ref{sec:evidence} since in that case we used an unbinned analysis to infer the presence of an additional component but not its magnitude. The large uncertainties in the profile reconstruction stem from the loss of information introduced by the binning procedure necessary to estimate the slope of the residuals. Despite this drawback, the profile reconstruction method constitutes a first step in the effort to extract the dark matter profile directly from rotation curve data without unnecessary adhoc assumptions, and its relevance at a precision level will grow as the quality of the available kinematic data improves throughout the decade.

\subsection{Modified Newtonian dynamics}\label{sec:mond}
\par The setup described in the previous sections can also be used to test and constrain modifications of gravity in the Milky Way. We have explored this direction in Ref.~\cite{Iocco:2015iia} in the context of modified Newtonian dynamics (MOND). Using two of the most popular interpolating functions (the so-called ``standard'' and ``simple'' functions), we have tested the MOND paradigm with the rotation curve data for each baryonic model, taking the critical acceleration $a_0$ as a fitting parameter. In the case of the standard interpolating function, our findings show that for most baryonic models MOND can explain the rotation curve of the Milky Way for values of $a_0$ significantly above those indicated by studies of external spiral galaxies (see e.g.~Ref.~\cite{2011A&A...527A..76G}). This friction is less prominent in the case of the simple interpolating function. While these results do not certainly rule out MOND, they confirm the existing tension between some MOND variants and observational data at the galactic scale.

\section{Future directions}\label{sec:future}

\par In this synopsis, we have summarised the considerable progress achieved by the community in constraining the dark matter density in the Galaxy over the decades and in recent years. A major conclusion is that -- whereas extensive photometric and kinematic data are indeed available -- the ignorance about the actual morphology of the visible component and current observational uncertainties in determining the Galactic fundamental parameters (i.e., the distance of the Sun to the Galactic centre and the local circular velocity) still are the dominating sources of uncertainty in pinpointing the dark matter component across the Milky Way. %This hinders searches for particle dark matter, especially those concerned with the innermost regions of the Galaxy. 
In the near future, a detailed determination of the dark matter density profile will depend crucially on our ability to shrink these two sources of uncertainty. On that respect, a new generation of astronomical surveys is steadily coming into place: the Gaia satellite \cite{2012Ap&SS.341...31D}, launched in December 2013 by ESA and already in its 5-year science data taking period since July 2014, the ground-based infrared survey APOGEE-2 (SDSS-IV, 2014--2020) \cite{apogee2site} covering both hemispheres, and the ground-based optical surveys WEAVE \cite{weavesite} (2017--2022) in the northern hemisphere and 4MOST \cite{2012SPIE.8446E..0TD} (2018--2023) in the sourthern hemisphere. These instruments will deliver a comprehensive, exquisite census of stars in the Galaxy, especially in the solar neighbourhood. The main challenge for the next decade is to use this batch of data to start a new precision era in mapping dark matter across the Milky Way.

\vspace{0.5cm}
{\it Acknowledgements.} M.~P.~acknowledges the support from Wenner-Gren Stiftelserna in Stockholm and F.~I. from the Simons Foundation and FAPESP process 2014/22985-1.

\bibliographystyle{JHEP.bst}
\bibliography{proceedingsICRC2015}

%\begin{thebibliography}{99}
%\bibitem{...} 
%....
%\end{thebibliography}

\end{document}